\documentstyle[pre,epsfig,aps]{revtex} 

\begin{document} 

\draft
 
\title{Propagating front in an excited granular layer} 
 
\author{W. Losert$^{1}$, D.G.W. Cooper$^{1}$, and J.P. Gollub$^{1,2}$} 
 
\address{$^{1}$Department of Physics, Haverford College, Haverford PA
19041, U.S.A.\\ 
$^{2}$Physics Department, University of Pennsylvania, Philadelphia 
PA 19104, U.S.A
}

\date{\today}

\maketitle

\begin{abstract}
A partial monolayer of $\sim 20000$ uniform spherical steel beads, 
vibrated vertically on a flat plate,
shows remarkable ordering transitions and cooperative 
behavior just below 1 g maximum acceleration.
We study the stability of a quiescent disordered or ``amorphous'' state 
formed when the acceleration is 
switched off in the excited ``gaseous'' state. 
The transition from the amorphous state back to
the gaseous state upon increasing the plate's acceleration
is generally subcritical:
An external perturbation applied to one bead 
initiates a propagating front that
produces a rapid transition. 
We measure the front velocity as a function of the applied acceleration.
This phenomenon is explained by a model based on a single vibrated particle 
with multiple attractors that is perturbed by collisions.  A simulation
shows that a sufficiently high rate of interparticle collisions can prevent 
trapping in the attractor corresponding to the nonmoving ground state.  
\end{abstract}
 

\section{INTRODUCTION} 

Granular materials exhibit phases that resemble yet are distinct from
ordinary solids, liquids, and gases. 
(For reviews see Refs~\cite{campbell90,jaeger96}.)
Granular materials at rest behave much like solids (e.g. they can 
sustain stress), but when
 excited they can behave as a dense liquid or a more dilute gas, 
with varying degrees of correlated motion.
Many phenomena observed in granular materials involve more than
one phase simultaneously. 
For example, in sheared granular materials and in avalanches both 
a solid phase and a liquid or gas phase are present. 
The coexistence of phases and the
mechanisms of transitions between phases are of great interest.
Although there are no attractive interparticle forces, it
is provocative to compare and contrast these phenomena to
their condensed matter analogs.

A critical difference is the presence of inelastic collisions
 and dissipation of 
energy in a granular medium~\cite{jaeger96}. 
Energy has to be added continuously in
order to maintain granular material in a fluid state. 
Dissipation can also render the excited state
 unstable against density fluctuations.
Regions of elevated density experience an elevated collision rate, 
which decreases average particle 
speeds and leads to further local density increases through the granular
equivalent of mass diffusion. 
The excited state collapses locally and 
clusters of particles at rest appear. This phenomenon has been studied 
numerically~\cite{McNamara92,Goldhirsch93,McNamara94,Luding98,Chen98}
 for the cooling of a granular gas without energy input~\cite{divcollisions}.
Clustering has been predicted to persist in the presence of 
excitation~\cite{Louge91},
whether the energy is provided homogeneously~\cite{Otha98,Vulpiani98}
or at the boundary~\cite{Kadanoff95}.
Clustering due to inelasticity has been studied experimentally by several
groups~\cite{Gollub97,Urbach98}.

Granular materials can be excited conveniently with a controlled energy input
through vertical vibrations of the container. 
Since energy is
transferred into horizontal motion through interparticle collisions,
the (mean) vertical kinetic energy is always larger 
than the horizontal kinetic energy\cite{knight96,warr952,delour98}.
Olafsen and Urbach \cite{Urbach98},  in a study of a vertically vibrated 
partial layer of steel spheres,
found spontaneous clustering into a two-dimensional ordered crystal at rest 
surrounded by a sea of vibrating particles, as the peak acceleration 
was reduced below 1g.
Both phases were found to coexist in steady state.

In this article, we study a 
triggered phase transition that transforms a quiescent disordered
phase to an excited granular gas via a propagating front.
Starting with a partial two-dimensional layer of spherical beads at rest 
on a vibrating flat plate
below a threshold vibration amplitude, 
we have observed that a small external perturbation of only
one bead can induce a phase transition into
a rapidly moving gaseous state that spreads
to all beads. 
We determine the frequency-dependent range of 
vibration amplitudes where the phase transition can be
triggered, and observe a subcritical-to-supercritical 
transition at a threshold frequency.
We also measure the growth rate and
interface shape of the gaseous region.
We account for our observations using numerical studies of a single 
periodically forced bouncing particle subjected to 
random perturbations arising from collisions.
The simulation explains the origin of the propagating front 
and may also help to explain the crystal-fluid coexistence
discovered earlier~\cite{Urbach98}.
 
Other metastable states are known in granular matter.
For example, a small perturbation suffices
 to trigger avalanches and a small (tapping) perturbation can
start the flow of granular material from a hopper. 
The propagating front described here is different from these examples;
it arises from the coexistence of two dynamical attractors.

\section{Experimental procedure and qualitative observations} 

The experiments are conducted in a rigid $32$~cm diameter
circular container made of anodized aluminum that is oscillated vertically
with a single frequency between $40$~Hz 
and $140$~Hz using a VTS500 vibrator from Vibration Test Systems Inc.
A computer controlled feedback loop keeps 
the vibration amplitude constant and reproducible.
Spherical (grade 100) nonmagnetic 316 stainless steel beads with 
$1.59 \pm 0.02$~mm diameter are
used as granular material.   
Most experiments were carried out with a single layer of $\sim 18400$ 
beads, which corresponds to a fractional coverage of $c=0.50$
(half that of a hexagonal close packed layer). 
The particles are illuminated from four sides and 
images are captured with a $512 \times 512$~pixel variable scan
CCD camera (CA-D2, Dalsa Inc.) at $2$~frames/s. 
In some experiments a $512 \times 480$~pixel fast CCD camera 
(SR-500, Kodak Inc.) operated at $30-60$~frames/s was used.
The coefficient of restitution for collisions with the aluminum plate
(i.e. the ratio of velocities after and before a collision), 
measured from images of successive 
vertical bounces taken with the fast camera, is
$\alpha_{\rm plate} = 0.95 \pm 0.02$ at commonly observed particle speeds. 
This result is similar to $\alpha_{\rm bead} = 0.93 \pm 0.02$, 
which was measured in Ref.~\cite{Gollub97}. 
(A similar value, $\alpha_{\rm bead} = 0.95 \pm 0.03$, was obtained
in Ref.~\cite{Louge97}.)
We found that $\alpha_{\rm plate}$ decreases with increasing impact velocity
(in agreement with Refs~\cite{Zippelius98,Dave97}). When rotational motion 
or a tangential velocity component is
present during collisions, the effective restitution coefficient decreases.

Conducting
nonmagnetic 316 stainless steel 
was chosen for the beads.
We found that the phase transitions occur at the same accelerations
when the shaker's  magnetic field was
reduced by $70\%$, and that the growth rate of the excited phase is unchanged.
These facts indicate that the external magnetic field 
generated by the shaker does not influence the
results presented here. We also did not find any significant electrostatic
effects.

\subsection{Quiescent disordered (amorphous) state}

\begin{figure}
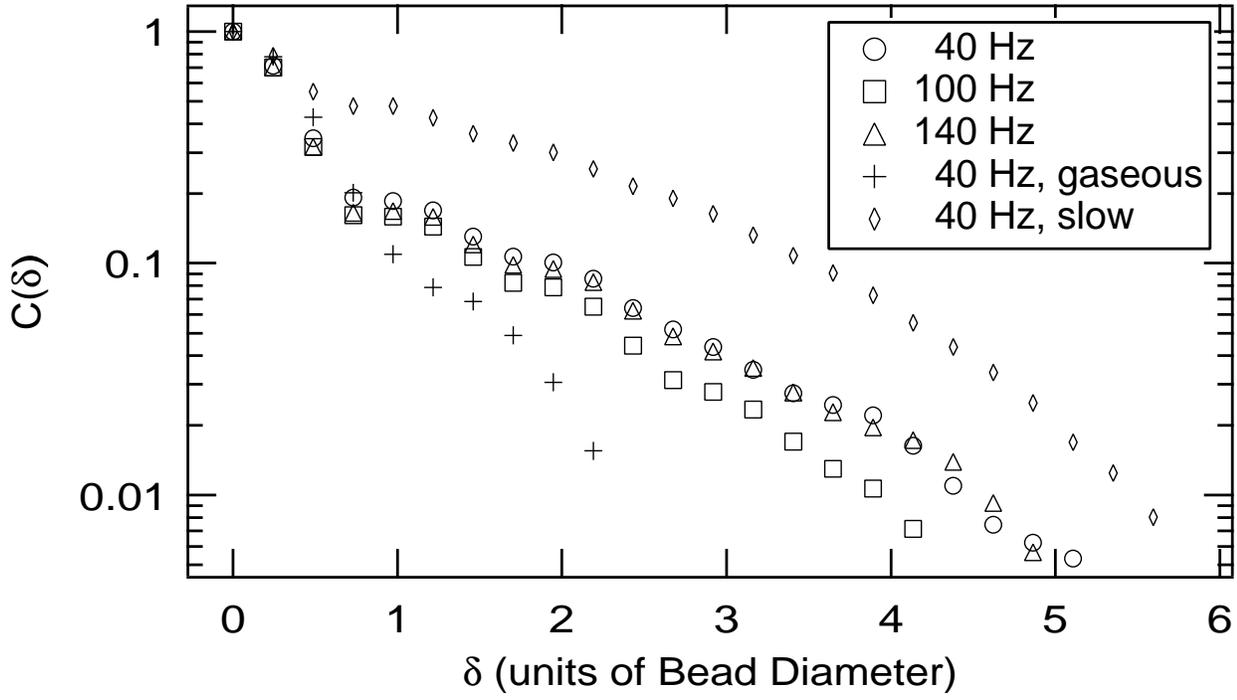

\caption{Radial autocorrelation function $C(\delta)$ 
of images of the amorphous state taken after
sudden cooling from initial states forced at different 
frequencies (triangles, circles, squares). 
Slow cooling (diamonds) produces stronger correlations.
Correlations are much weaker in the gaseous state(crosses).} 
\label{corr_amorph} 
\end{figure}

In order to create reproducible initial conditions 
for all experiments, the container is vibrated for 
more than $20$~s at a peak acceleration $a \ge 2$~g. 
Then the vibration is abruptly turned off and the particles
come to rest through free cooling (i.e. decrease of
particle energies through inelastic collisions).
Figure~\ref{corr_amorph} shows the resulting radial autocorrelation function 
$C(\delta)$ of images of the amorphous state
after sudden cooling starting from different forcing frequencies;
the horizontal axis is scaled by the bead diameter.
The correlation function is calculated from the
greyscale intensity $I$ after subtraction of the average 
greyscale intensity as 
$C(\delta)=\langle I(r) I(r+\delta) {\rangle}_r / 2 \pi \delta 
\langle I(r)^2{\rangle}_r$.
We note a sharp peak near $\delta=0$ corresponding to the image of a 
single bead, and
a tail that is insensitive to the vibration frequency and 
amplitude prior to the shut off.
This tail is not present in the gaseous state, also shown.
The clustering of beads during
sudden cooling, which was anticipated numerically, appears to be
insensitive to the frequency at which the beads were excited. 
When the amplitude is lowered slowly, clustering becomes
measurably stronger, as is evident from the upper curve.

\subsection{Propagating front}

\begin{figure}
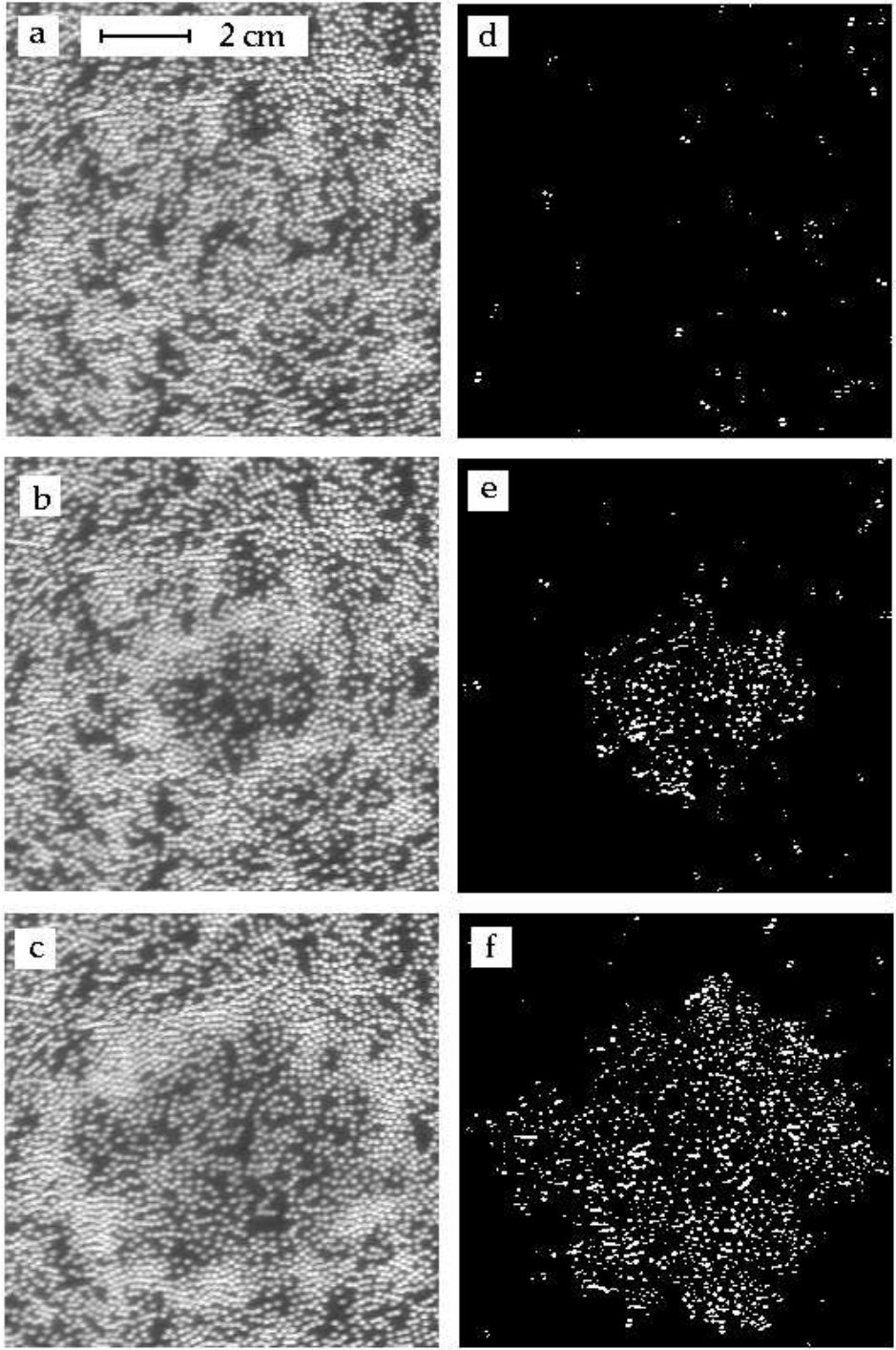
 
\caption{Front propagation during the amorphous-to-gaseous 
transition at $80$~Hz with $a=0.93$~g. Images captured every 
$3.5$~s are shown at left; absolute differences between images taken 
$0.5$~s apart are shown on the right to
highlight moving particles. (a,d) Amorphous initial state; 
(b,e),(c,f) unstable front propagation.} 
\label{melt_img} 
\end{figure}

The focus of the experiments presented in this article
is the transition to a moving gaseous state, 
starting from the reproducible quiescent amorphous state 
produced by sudden cooling.
If the peak acceleration is much larger than 1~g, all beads start
moving  immediately when the vibration is switched on.
For peak accelerations much smaller than 1~g, the beads remain quiescent on
the plate indefinitely and return into the quiescent state when perturbed 
externally. 
In contrast, for a frequency dependent intermediate range of peak 
accelerations just below 1~g, the beads remain at rest
when the vibration is switched on, but a transition to the moving gaseous 
state can be triggered by setting at least one bead into motion. 
Figure~\ref{melt_img} shows the typical evolution of a portion of the layer 
from the amorphous state to the gaseous state when a region of several beads
is perturbed.
The peak acceleration $a$ of the vibrated plate
has been set below, but close to, a threshold $a_c$
 that is dependent on driving frequency $f$; 
the beads remain essentially at rest on the plate
(Fig~\ref{melt_img}a).  
An external perturbation is then applied to a few beads, either by rolling
an additional bead through an inclined cylinder onto the surface or
by pushing a few beads manually.
The perturbed bead(s) start to bounce and
move in the horizontal direction, transmitting energy to their 
neighbors, which in turn start to move. 
An area within which all beads are moving develops quickly, surrounded 
by a denser area that marks the interface between the granular gas and
 the essentially static amorphous state (Fig.~\ref{melt_img}b).
The area of the moving gaseous phase increases rapidly
as the dense front propagates
into the amorphous phase (Fig.~\ref{melt_img}c), 
until all beads are moving. 
For accelerations where a crystalline phase can exist (see below),
we observe the development of a crystal during the transition
and its coexistence with the gaseous phase in the steady state.

\subsection{Interface shape}

\begin{figure}
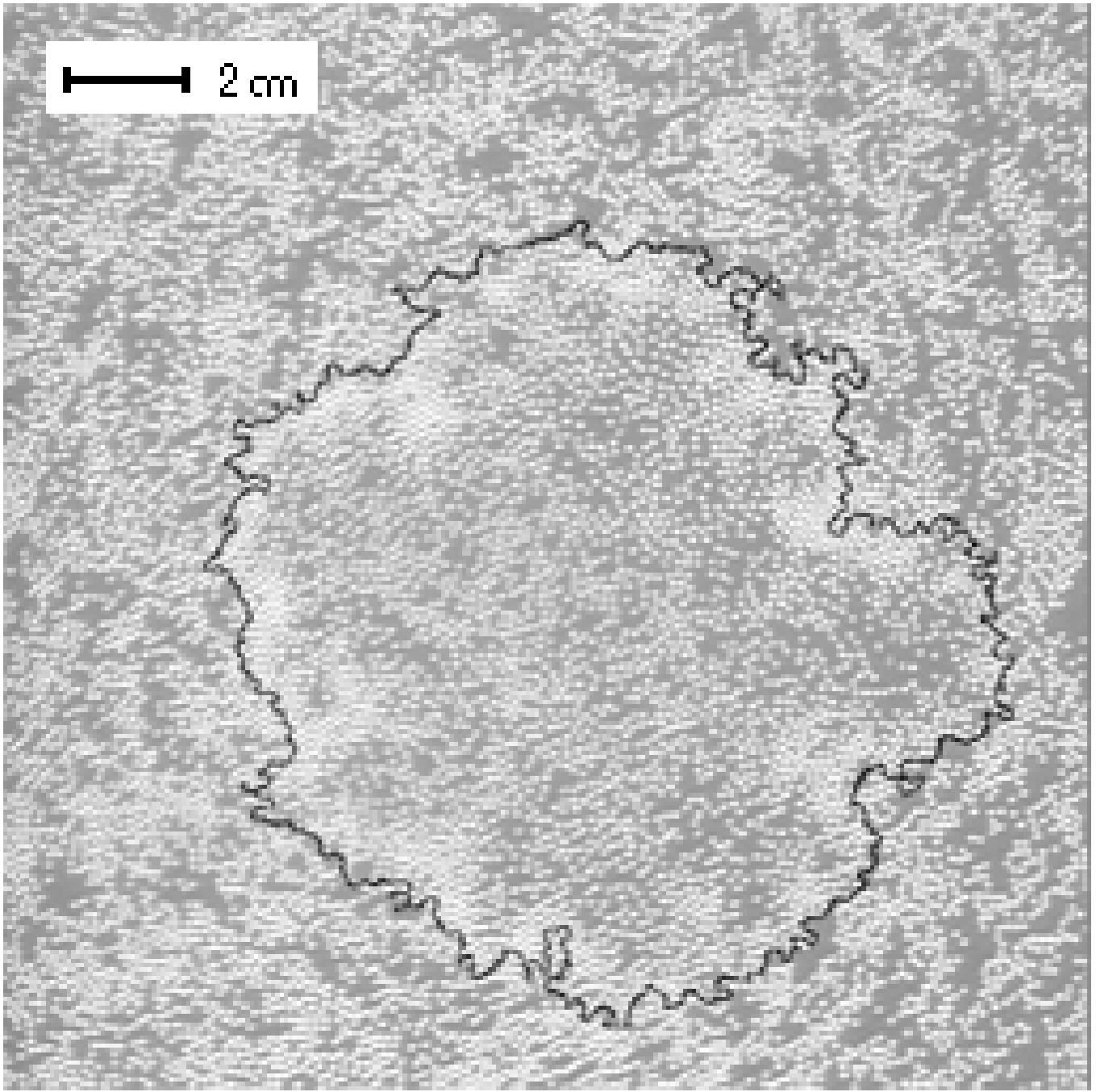
 
\caption{Perimeter of the gaseous region (black line) superimposed on 
an image of the beads. 
The dense front clearly moves as the gaseous region expands.} 
\label{overlay} 
\end{figure}

We use the absolute difference of two consecutive frames to 
calculate the area and the interface shape of the gaseous region.  
Such a difference image, converted to black and white,
 produces a white spot at the initial and final positions of a moving bead,
 leaving all stationary beads and the background black.  
This allows us to distinguish between the
 essentially stationary amorphous phase and the moving gaseous phase.  
Figures~\ref{melt_img}d-f are
 difference images corresponding to Fig.~\ref{melt_img}a-c.  
(The difference images reveal that a few beads do move even in 
the amorphous phase, so small spontaneous 
perturbations are always present.)
In order to extract the 
size and shape of the gaseous region from the difference images,
we create a continuous white region using NIH image. The perimeter of
the gaseous region, extracted with this method from the absolute 
difference of two consecutive frames, is shown overlayed onto 
the second frame in Fig.~\ref{overlay}. The dense front clearly moves as
 the gaseous phase expands, so 
we consider the dense front as part of the gaseous 
phase~\cite{foot1}.

\section{QUANTITATIVE RESULTS} 

\subsection{Phase Diagram}

\begin{figure}
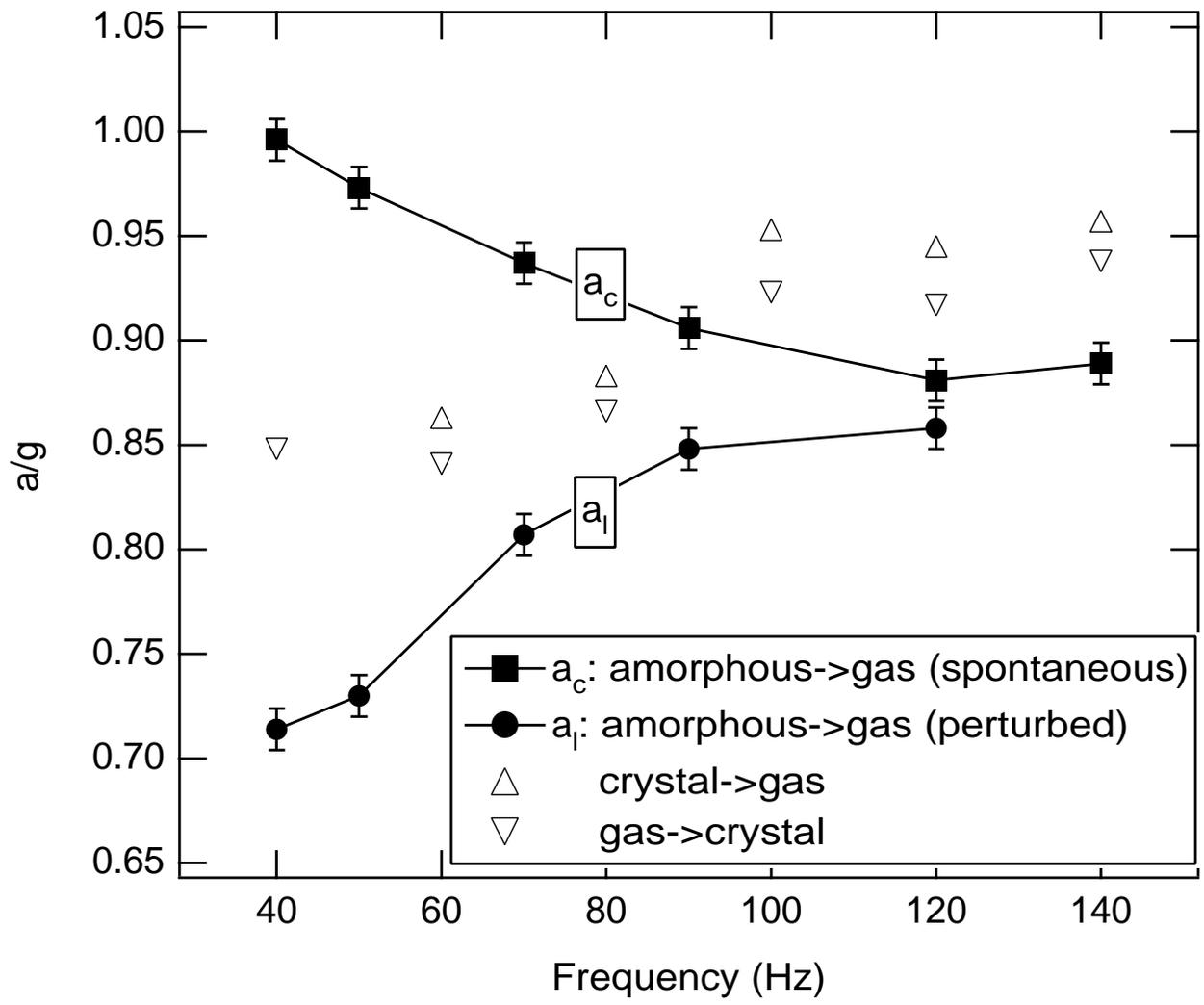

\caption{Hysteresis of the amorphous-to-gaseous phase transition. 
Peak accelerations (normalized by g) for perturbed and spontaneous 
amorphous-to-gaseous transitions are shown (closed symbols). 
The freezing and evaporation points of the crystalline phase 
are also shown (open symbols).} 
\label{meltpnt} 
\end{figure}

The transition between
the  amorphous and gaseous phases, triggered by external perturbations, 
 can occur only for a small range of vibrator accelerations.
Below a low acceleration limit $a_l$, no perturbation is able to initiate 
a gaseous region that persists for at least one minute.
Above a higher acceleration $a_c$, 
the spontaneous perturbations present in the amorphous state 
(observable in Fig.~\ref{melt_img}d), are sufficiently strong
to trigger a propagating front within two minutes
without an external perturbation. 
Figure~\ref{meltpnt} shows  $a_l$ and $a_c$  
as a function of the driving frequency $f$. 
The hysteretic region between $a_c$ and $a_l$ 
was found to be reproducible to
within $1\% $ of the driving acceleration.
It decreases
with increasing frequency, and a subcritical to supercritical transition 
takes place at $f_t \approx 120$~Hz. At higher frequencies, 
spontaneous excitation of the gaseous phase is observed for  $a \le a_c$, 
but any excitation decays slowly, 
and even large external perturbations fail to
 trigger a transition into the gaseous state.
For $a > a_c$, on the other hand, areas of local excitation develop 
(often in numerous spots simultaneously) and propagate quickly 
throughout the system. 
The freezing and evaporation points of the crystalline phase
are also shown in Fig.~\ref{meltpnt} as open symbols. 
At lower frequencies they fall in the middle
of the hysteretic region, while at higher frequencies 
(roughly above $f_t$) they lie above the amorphous-to-gaseous transition.  
When the freezing point falls within the hysteretic region, 
the amorphous-to-gaseous transition leads to a gaseous phase for accelerations
above the freezing point, and to coexisting crystalline and gaseous
regions at accelerations
below the freezing point.
Additional measurements indicate that 
at a lower bead coverage $c=0.25$, 
both $a_l$ and $a_c$ change by less than $5\%$.

\subsection{Growth Rates}

\begin{figure}
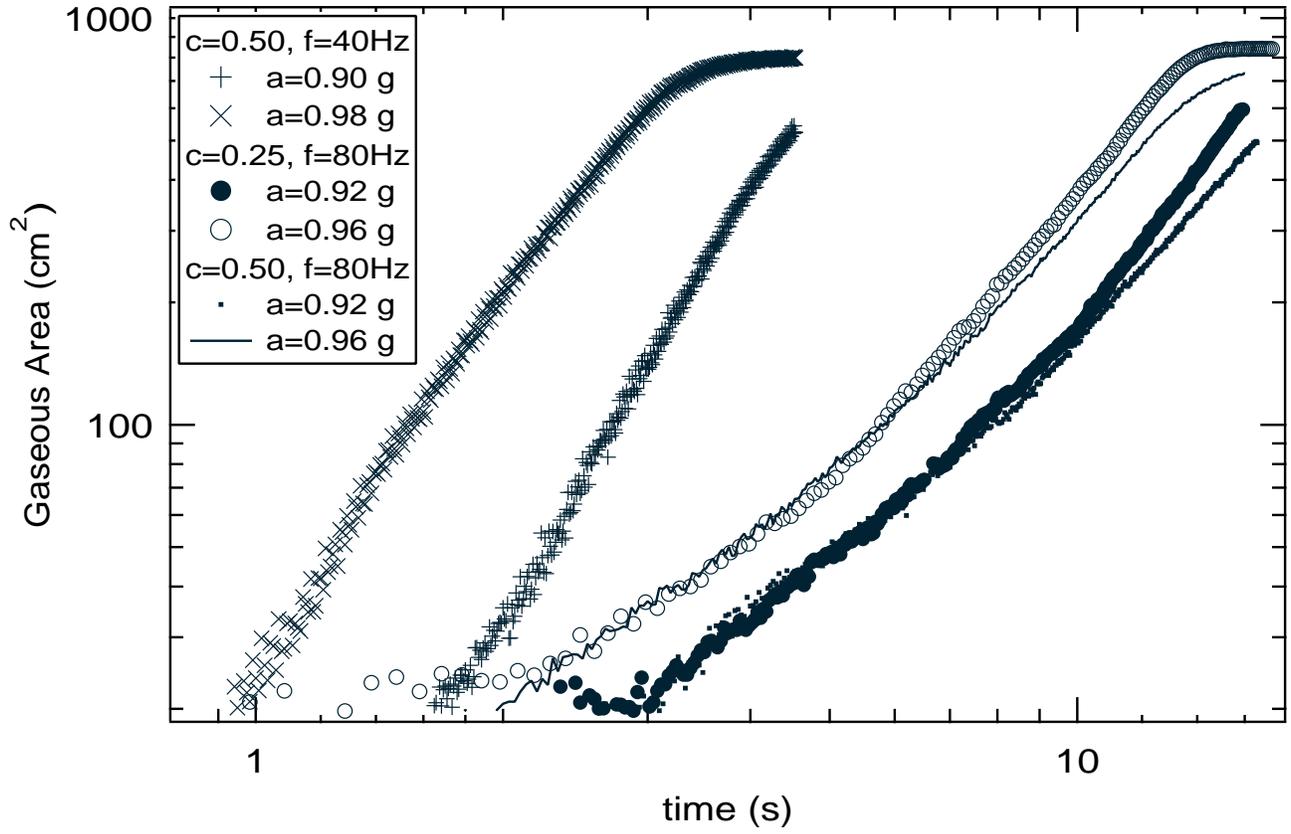
 
\caption{Growth of the gaseous area $A$ for various peak accelerations,
bead coverages, and vibration frequencies.
Smaller frequencies and larger accelerations lead to faster melting. 
Lower coverage leads to faster melting for large areas only.} 
\label{meltrate} 
\end{figure}

The growth rate of the gaseous 
region can be measured between $a_l$ and $a_c$ (and even somewhat
 above $a_c$ by applying an external
 perturbation quickly, before a spontaneous instability occurs).
Figure~\ref{meltrate} shows  a double logarithmic plot of  
the area $A$ of the gaseous region
as a function of time after initiation by a perturbation.
The measurements are carried out at 
$40$~Hz and $80$~Hz for several peak accelerations within the hysteresis
region and slightly above $a_c$. 
A doubling of the driving frequency slows the rate of growth 
of the gaseous area 
by approximately a factor of 4. Reducing the coverage does not change the 
initial growth rate, but leads to faster growth at later times. The total
plate area $A_{\rm max}=804~ {\rm cm^2}$ limits the growth eventually.
Figure~\ref{meltrate} indicates that the growth, after an initial transient
during which the dense front forms, can be 
approximately described by a power law with an exponent that is
independent of acceleration and only slightly dependent on coverage.
However, the exponent changes significantly with frequency.  

A realistic description of the
 dynamics of growth of the gaseous region is complicated by the
dynamics of the dense layer of beads ahead of the interface,
which appears to slow the growth of the 
gaseous region. The dense layer is pushed uniformly by the advancing 
gas under some conditions (see Fig.~\ref{overlay}), 
but in other cases crystalline regions form that do 
not move. The advancing front then grows around those regions, 
leaving behind slowly melting crystalline regions within the gaseous phase.    
At small coverages ($c \sim 0.05$) an interface between
the amorphous and gaseous phase is not defined, as moving beads can pass many 
stationary beads between collisions. In this case the rate of increase in the 
number of moving beads is only limited by the mean interval between collisions 
with stationary beads.

\begin{figure}
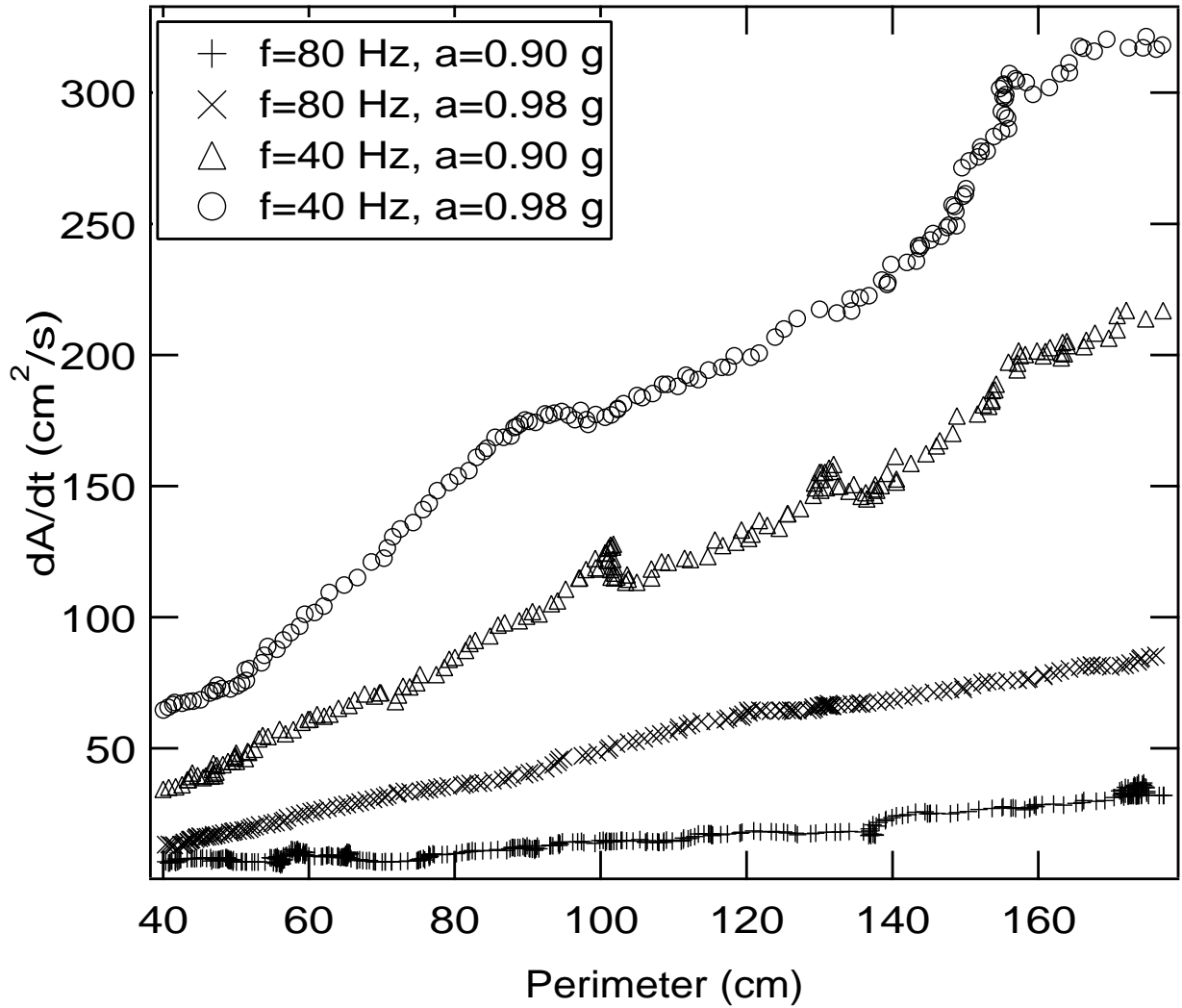
 
\caption{Growth rate of the gaseous area $dA/dt$ vs. perimeter length $P(A)$
for several peak accelerations and vibration frequencies.
The dependence is approximately linear.
Both $dA/dt$ and $P(A)$ are smoothed by means of 
a running average over 11 points.} 
\label{dadtvsp} 
\end{figure}

\begin{figure}
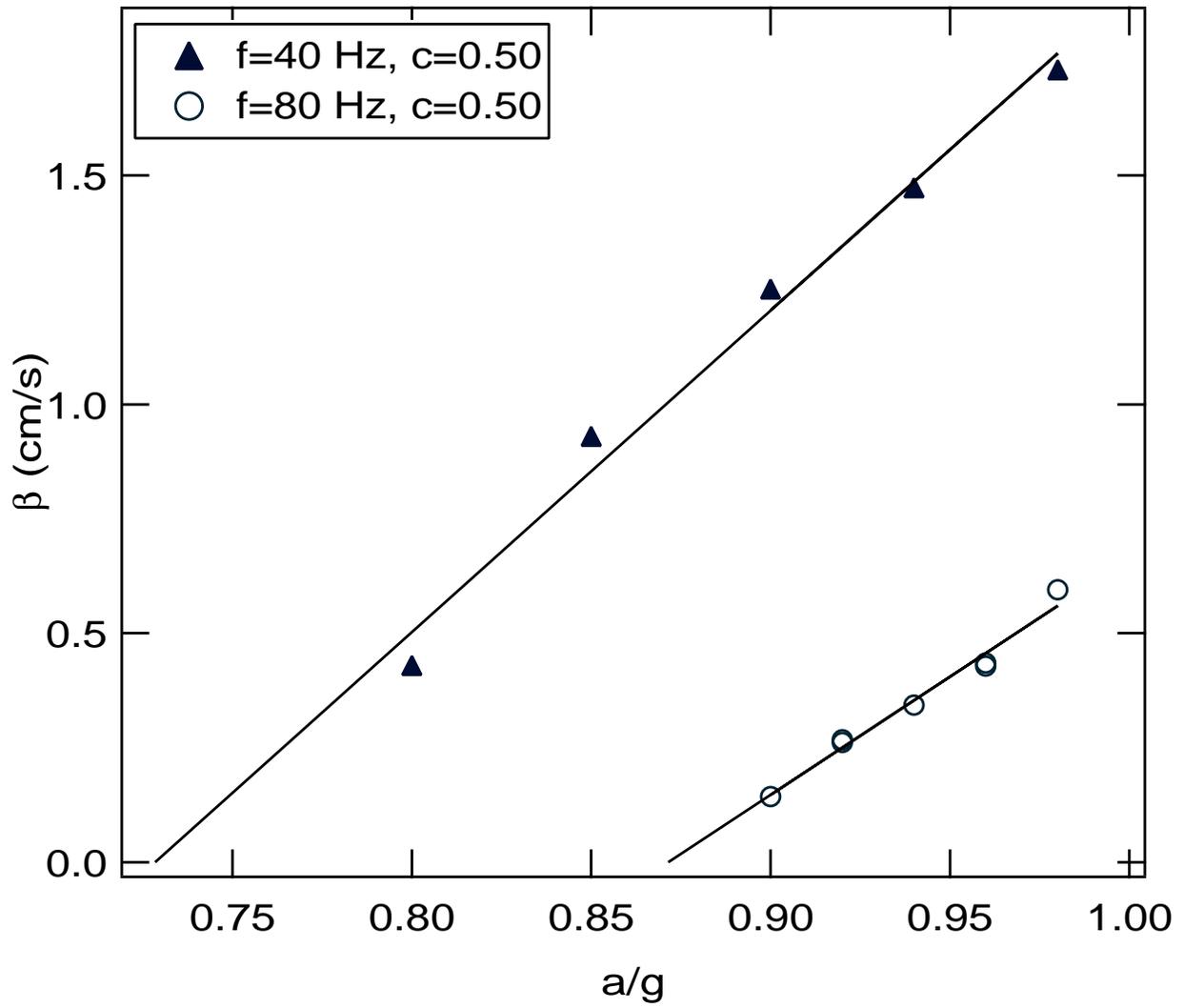
 
\caption{Front velocity $\beta$ vs. peak plate acceleration
(normalized by g) for different
frequencies.
The dependence is linear; $\beta \to 0$ near $a_l$.} 
\label{growth_rate} 
\end{figure}

It is nevertheless possible to describe the growth process approximately
 by noting that the growth rate $dA/dt$ increases 
in proportion to the perimeter $P(A)$ between the gaseous 
and the amorphous regions as shown in Fig.~\ref{dadtvsp}
for areas between $\sim 50$ and $400~{\rm cm^2}$. We find
\begin{equation}
dA/dt = \beta (a,f) P(A),
\end{equation}
where the front velocity  $\beta$ depends on $a$ and $f$. 
This indicates that the growth rate of the  gaseous area
 is roughly proportional to the rate at which beads collide with
 the interface once a dense layer has formed. 
The front velocity $\beta$ is shown in Fig. \ref{growth_rate}. 
It increases approximately linearly with $a$
and goes through zero near $a_l(f)$.

\section{BOUNCING BALL MODEL} 

\subsection{Model of unperturbed bouncing}

The hysteretic nature of both the amorphous-to-gaseous transition (and 
to a lesser degree the crystalline-to-gaseous transition \cite{Urbach98})
 below $1{\rm ~g}$ peak acceleration
indicates that an energy gap exists between the states at rest 
(amorphous and crystalline) and
the lowest energy excited steady state.
The simplest model that can be used to explain the observed behavior 
is based on the dynamics of a single bead 
bouncing on a plate under sinusoidal vibrations
$x_{p}=X_{\rm max} cos(2 \pi f t)$.
The peak acceleration then is $a = X_{\rm max} 4 \pi^2 f^2$.
In a collision with the plate, the vertical bead velocity changes from
$v$ to $v^{'}$:
\begin{equation}
v^{'} = (1+\alpha_{\rm plate})\dot{x}_{p} - \alpha_{\rm plate} v.
\label{eq1}
\end{equation}
The bead at rest on 
the plate ($v^{'} = v = \dot{x}_{p}$)  always represents
a stable steady state solution below $1$~g peak acceleration
with an average bead energy
\begin{equation}
E_0=\frac{a^2}{16 \pi^2 f^2}
\end{equation}
per unit mass.
Additional steady states appear as the acceleration is increased. 
The first nonsticking steady state is a periodic bouncing with speed
at the instant of collision given by:
\begin{equation}
v^{'} = -v = \frac{1+\alpha_{\rm plate}}{1-\alpha_{\rm plate}} \dot{x}_{p}.
\label{eq2}
\end{equation}
For periodic bouncing every vibration period this requires: 
$v^{'} = g / (2 f)$ and therefore (at the moment of contact) 
\begin{equation}
E_1=\frac{g^2}{8f^2} 
\label{eq3}
\end{equation}
 per unit mass.

Other excited states that are observable under conditions similar to the 
experiment have higher energy.  
States with a period $n/f$, where the bead hits the plate on average
once per vibration period and the impact pattern is repeated after $n$ 
cycles, have an average energy of $E=(g^2/8n) 
\sum_k \delta t_k^2$.  The minimum of $E$ occurs for 
the simple periodic state $n=1$, where 
the time intervals between contacts with the plate $\delta t_k$ are of 
equal length.
Periodic or chaotic states with more than one plate impact per
vibration period would have lower energy if they exist, 
but no such state was found in several calculations of basins of attraction
using the bouncing ball program by Tufillaro {\it et al}~\cite{Tufillaro92}
with parameters similar to the experiment. 
Those states therefore either do not exist, are unstable, or have
a negligible basin of attraction.

For vibrator accelerations below $1 {\rm ~g}$ 
an energy gap therefore exists between the lowest energy steady state $E_0$,
which is the quiescent state, and the first excited state,
a periodic bouncing state with energy $E_1$. 
We propose that this energy gap determines the size of the hysteretic region.
The measured decrease in the extent of hysteresis with increasing $f$ 
may reflect the dependence of $E_1$ and $E_0$ on $1/f^2$.

The discussion so far, which is based on a one dimensional idealized model, 
 has neglected perturbations. In the experiments,
however, perturbations are important, as each bead is subjected
to frequent interparticle collisions (which create
and sustain horizontal motion) and to other perturbations 
(e.g. due to slight nonuniformities of the vibrator surface). 
Since a perturbation can trigger a transition into a different steady state, 
we include them in a model that can be 
simulated numerically. 

\subsection{Simulations}

We follow the vertical movement of a single bead that
is subjected to random perturbations of its vertical speed 
representing
 collisions with other beads. 
Our strategy is to determine the mean energy of the 
vertical motion and to compare it with that of the first excited state.  
Therefore, we do not track the horizontal speed of the particle.
The time of perturbation events is selected randomly to create
an average perturbation rate $f_p$. The velocity of
the bead after a perturbation is $v^{'} = v \alpha_{\rm bead} r$,
where $r$ is a random number with gaussian distribution and unit variance.
To verify that the results do not depend on the choice of the 
probability distribution, we also tried letting $r$ be either $\pm 1$
(randomly), with similar results.

Energy losses from interparticle collisions are included 
through $\alpha_{\rm bead}$.
Bead and plate positions are calculated at fixed
 time intervals  $\delta t$. 
The time of bead impact on
the plate is calculated to within $O(\delta t^{2})$. 
The bead is considered to be stuck if 
more than one bounce per time interval occurs.
The coefficients used in the simulations were $f=80$~Hz,  
$\delta t = 5 \times 10^{-5} {\rm ~s}$,
  and $\alpha_{\rm plate}=\alpha_{\rm bead}=0.90$ (a rough estimate for the
effective coefficient of restitution in the presence of 
bead rotations during impact).
After an initial transient time of $25$~s 
the average behavior of the bead is recorded for $475$~s.

Since the calculations are carried out for one bead only,
it is not determined a priori 
whether the perturbation rate is sustainable from interbead collisions
in a many particle system.
Many particles must be in an excited steady state to sustain 
continued collisions, since two quiescent particles cannot collide.  
Collisions are only able to maintain or increase the number of excited 
particles (a necessary condition for sustainability of the collision rate)  
if the steady state mean energy is comparable to $E_1$.
Significantly lower computed mean energies would actually 
not be sustainable in the many particle system, 
because collisions would lead to a loss of particles from 
the excited state, and the system would settle into the ground state 
attractor for all particles.  

To check the conditions under which $E\ge E_1$, we show in 
Fig.~\ref{energy_sim} the average energy of the particle 
as a function of $f_p$. 
As the perturbation rate is increased, the energy of 
the particle first increases and then decreases. 
For comparison, the energies of a single
bead in the lowest excited state $E_1$ and in the quiescent state $E_0$
of the unperturbed problem are
shown as horizontal lines.
At $a=0.66$~g, $E<E_1$ for all 
perturbation rates;
therefore the particle will be trapped in the stationary ground state.
For $a > 1 {\rm ~g}$, $E>E_1$  the mean particle energy is larger
than $E_1$ (except for $f_p/f >>1$); 
here the quiescent state attractor does not exist.
On the other hand, for $0.8$~g~$ < a < 1$~g, $E>E_1$ for 
a limited range of $f_p$; this regime will show hysteresis. 

This explains qualitatively the occurrence of hysteresis for a many
particle system that is sufficiently dense to provide the required collision
rate. The propagating front is basically a self-sustained chain reaction.

\section{DISCUSSION AND CONCLUSIONS} 

In this paper we have described the steady states 
and phase transitions of a two dimensional
layer of beads subjected to vertical vibrations below $1{\rm ~g}$ 
peak acceleration. We start from a nearly static
amorphous state created by temporarily stopping
the vibration while all particles are moving rapidly.
Free cooling of the granular layer
creates a reproducible structure, with a correlation function
that is not dependent on the driving frequency or acceleration
over a wide range.

Starting from this quiescent disordered (``amorphous'') state,
we observe a striking transition to the gaseoue state, via a 
propagating front, as the vibration amplitude is increased.
A region of hysteresis exists for which the amorphous phase remains
metastable but a perturbation induces a transition.
The final state can consist either of gas, or of crystal and gas, depending 
on the acceleration and coverage.
The velocity of the propagating 
front $\beta$, which remains constant during the transition,
 increases linearly with peak plate acceleration $a$ and decreases 
with increasing vibration frequency $f$.
 
A simple model, based on a single bouncing bead with random 
perturbations modeling
collisions, exhibits behavior that is 
consistent with the experimental results. 
An energy gap between the ground and excited states of the single bouncing 
bead leads to a chain reaction that can sustain the excited
state of the many-particle system under the right conditions. 
The phenomenon is similar in some 
respects to a flame front.

The coexistence of the ground state and excited state attractor
can be seen most clearly at low coverage, where
collisions become sufficiently 
rare that beads can remain in the ground state attractor long enough 
to become stationary.
In Fig.~\ref{speed_bounce} we show a time average image at $c=0.03$.  
Moving beads appear in this figure as streaks 
and stationary ones as bright spots.
In addition, small clusters of beads at rest are clearly observable. 

The crystal-gas coexistence
is another potential manifestation of the coexistence of the
two attractors, in this case spatially separated by an interface. 
The simulation therefore suggests a possible origin
of the crystalline phase discovered by Olafsen and Urbach~\cite{Urbach98}. 
As the acceleration is lowered, 
the maximum energy provided by the plate
decreases until more energy is
lost in collisions than can be supplied by the plate. 
Beads then come to rest in the dense regions of the plate where
the collision losses are greatest. A crystal forms as the moving 
beads apply pressure to the cluster at rest, forcing it into
the densest possible configuration. Bouncing beads that hit the crystal
tend to come to rest at the interface, increasing the size of the
crystal but simultaneously decreasing the bead concentration
in the remaining gaseous phase. Crystal growth stops
when the density of the gaseous phase is lowered 
sufficiently that the 
(density-dependent) collision rate can be sustained in steady state.

We conclude that the rich variety of steady states that are observed just
below $1$~g peak vibrator acceleration are related to the coexistence of a 
ground state attractor 
that exists up to $1$~g 
and excited state attractors that
exist below $1$~g. The rapidly propagating front of highly excited
beads spreading in a sea of beads at rest,
which can be triggered by one moving bead without a change in the 
acceleration amplitude,  
is a striking many body consequence of this coexistence.

\section{Acknowledgments} 

This research was supported by the National Science Foundation
under grant No. DMR-9704301. We appreciate helpful discussions with 
I. Aronson, J.-C. Geminard and J. Urbach.


\begin{figure}
\caption{Simulation: Average energy of a bead vs. perturbation frequency, and
comparison to energy of the lowest excited state $E_1$ and quiescent state
$E_0$ (horizontal lines). 
Above a threshold acceleration $a \approx 0.8$~g
the mean energy is larger than $E_1$ for a range of perturbation
frequencies, a necessary condition for maintenance of the excited state in 
the many particle system.} 
\label{energy_sim} 
\end{figure} 

\begin{figure}
\caption{Coexistence of beads at rest and moving beads in a steady state
at low coverage.
Ten successive images, taken at $30$~frames/s were averaged
($c \approx 0.03$; $a=0.94 $~g; $f=80$~Hz).
Moving beads appear as streaks and stationary beads as bright spots.
Small clusters of beads at rest are clearly observable.} 
\label{speed_bounce} 
\end{figure} 

\setcounter{figure}{0}
 
\pagebreak

\begin{figure}
\begin{center} 
\epsfig{file=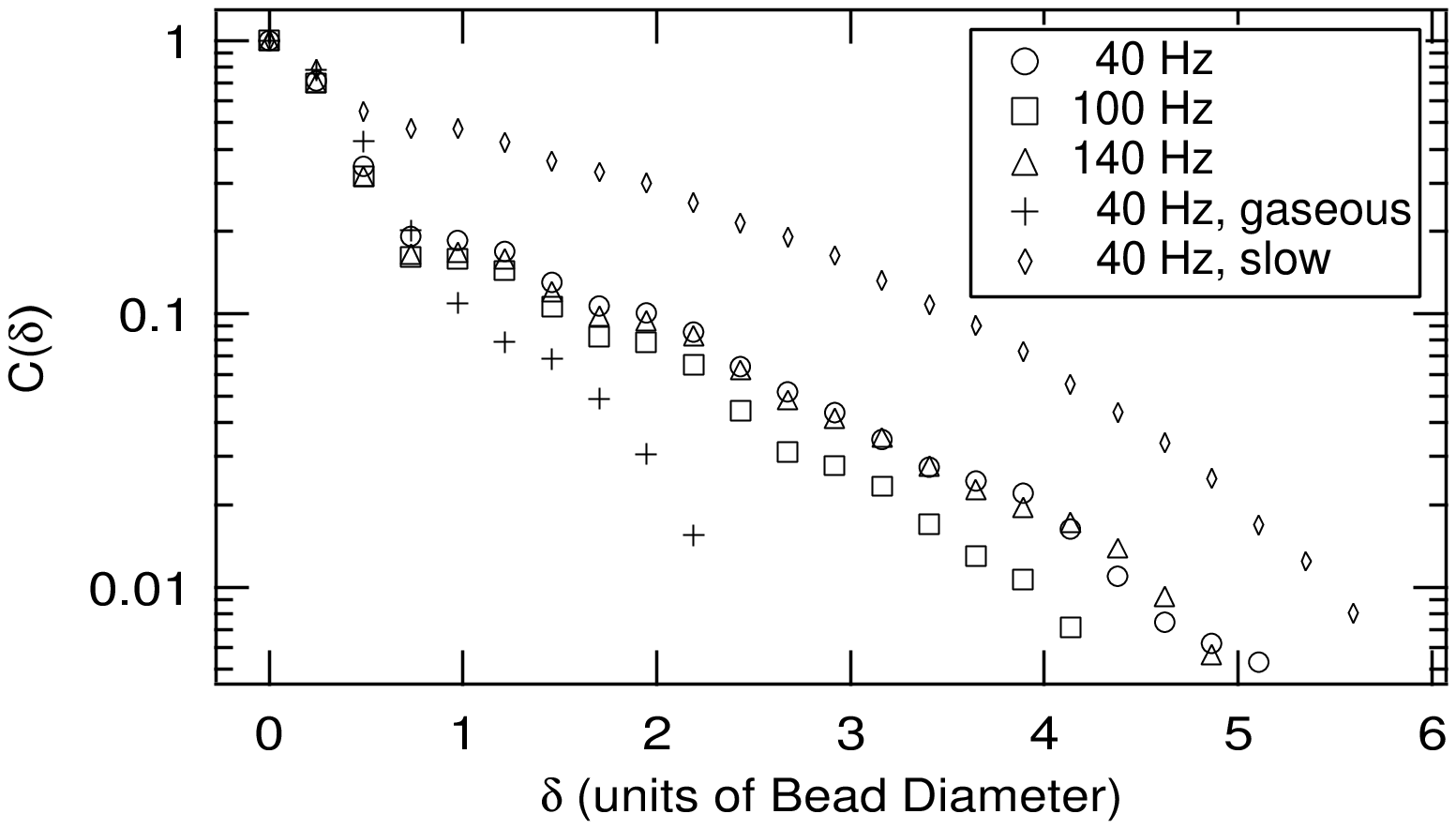, width=\linewidth} 
\end{center}
\caption{Radial autocorrelation function $C(\delta)$ 
of images of the amorphous state taken after
sudden cooling from initial states forced at different 
frequencies (triangles, circles, squares). 
Slow cooling (diamonds) produces stronger correlations.
Correlations are much weaker in the gaseous state(crosses).} 
\end{figure} 
\pagebreak

\begin{figure} 
\begin{center} 
\epsfig{file=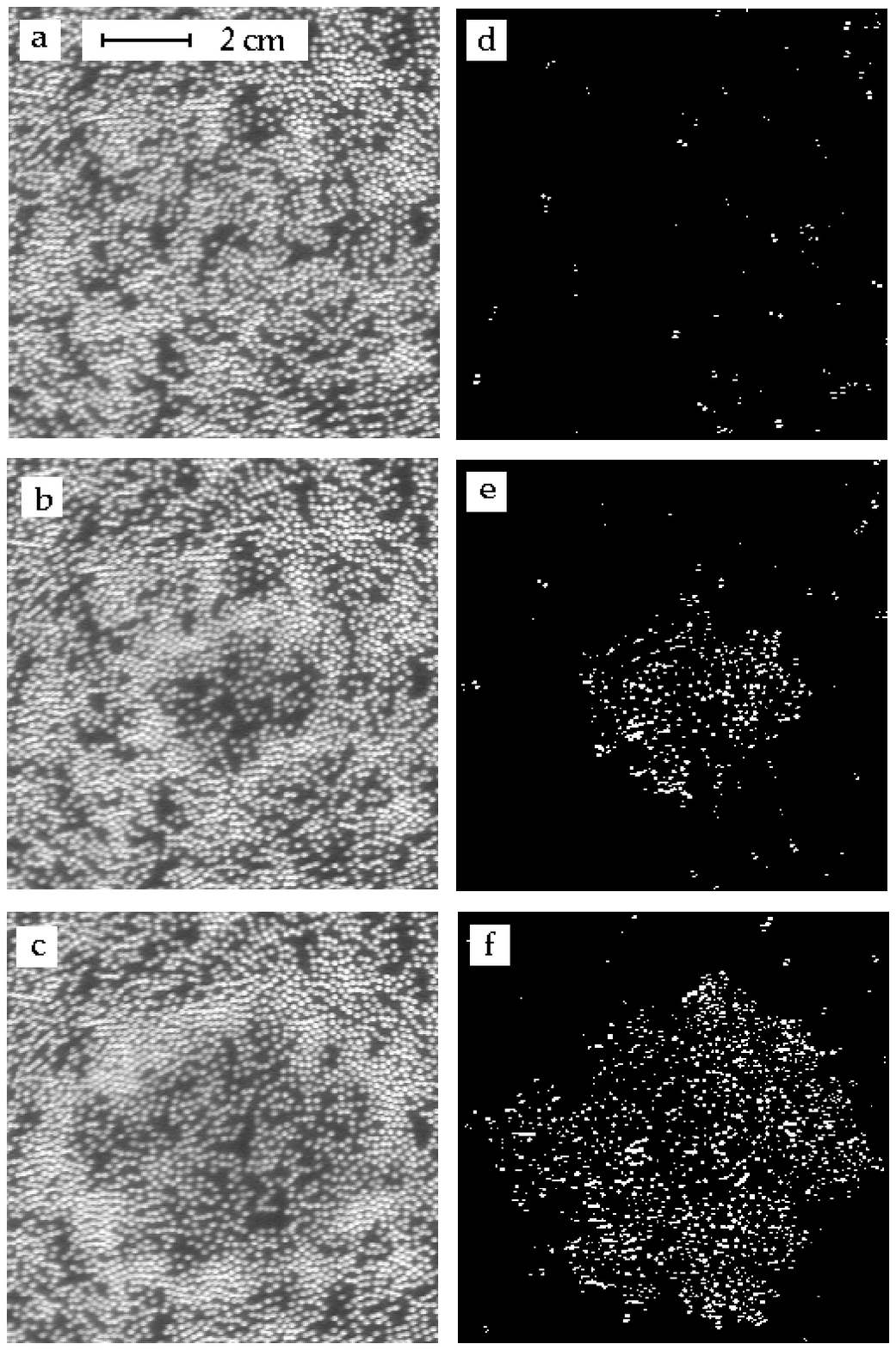, width=13 cm} 
\end{center}
\caption{Front propagation during the amorphous-to-gaseous 
transition at $80$~Hz with $a=0.93$~g. Images captured every 
$3.5$~s are shown at left; absolute differences between images taken 
$0.5$~s apart are shown on the right to
highlight moving particles. (a,d) Amorphous initial state; 
(b,e),(c,f) unstable front propagation.} 
\end{figure} 
\pagebreak
 
\begin{figure} 
\begin{center} 
\epsfig{file=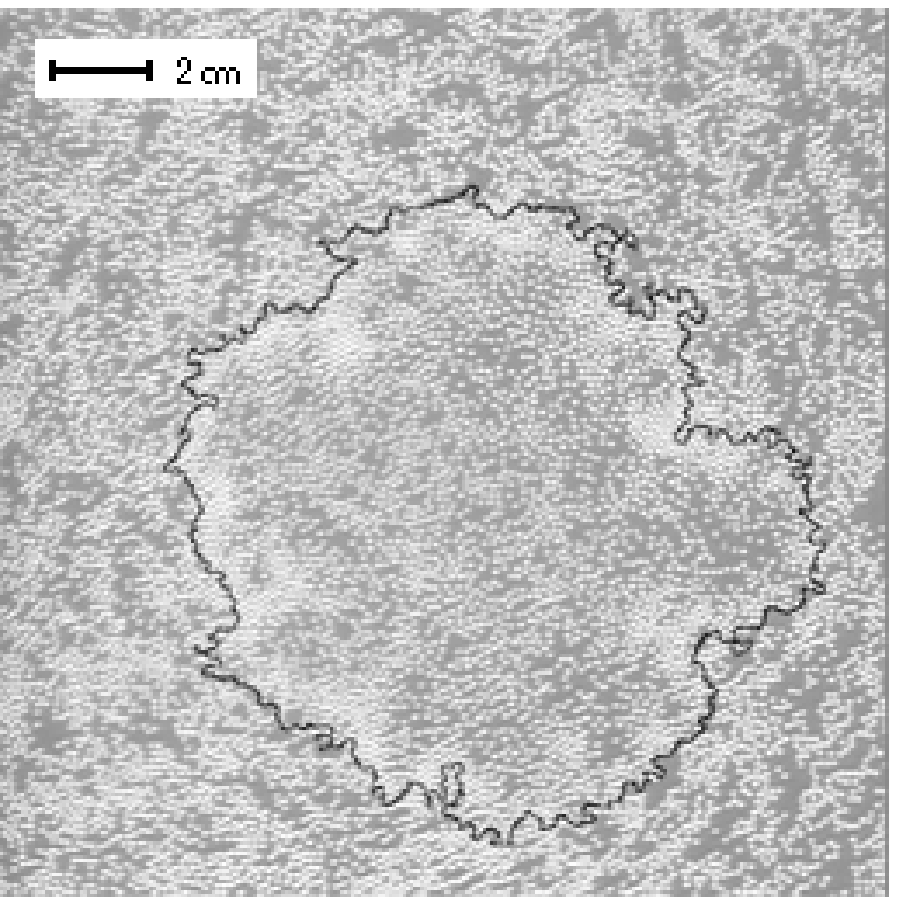, width=\linewidth} 
\end{center}
\caption{Perimeter of the gaseous region (black line) superimposed on 
an image of the beads. 
The dense front clearly moves as the gaseous region expands.} 
\end{figure} 

\begin{figure}
\begin{center} 
\epsfig{file=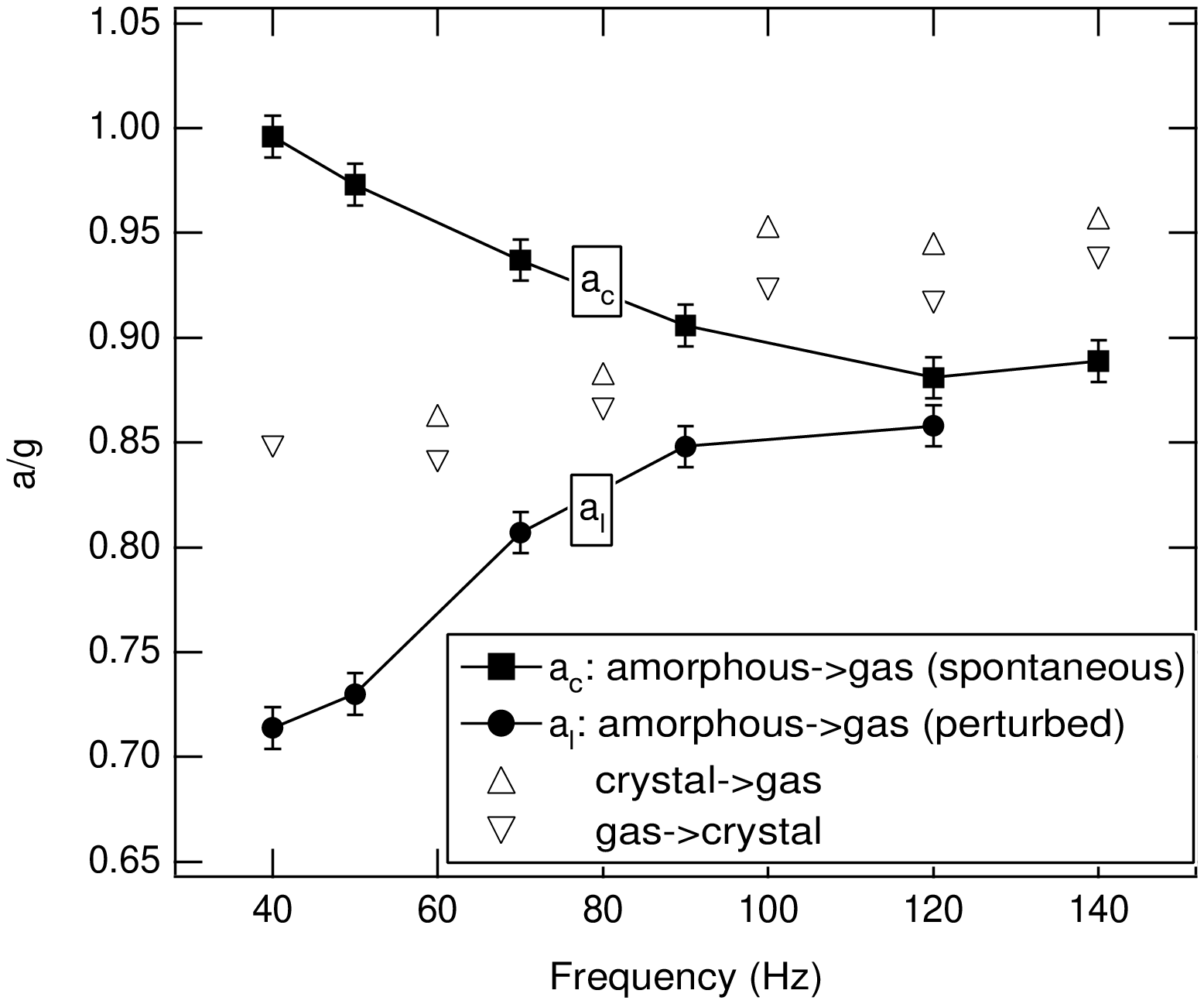, width=\linewidth} 
\end{center}
\caption{Hysteresis of the amorphous-to-gaseous phase transition. 
Peak accelerations (normalized by g) for perturbed and spontaneous 
amorphous-to-gaseous transitions are shown (closed symbols). 
The freezing and evaporation points of the crystalline phase 
are also shown (open symbols).} 
\end{figure} 

\begin{figure*} 
\begin{center} 
\epsfig{file=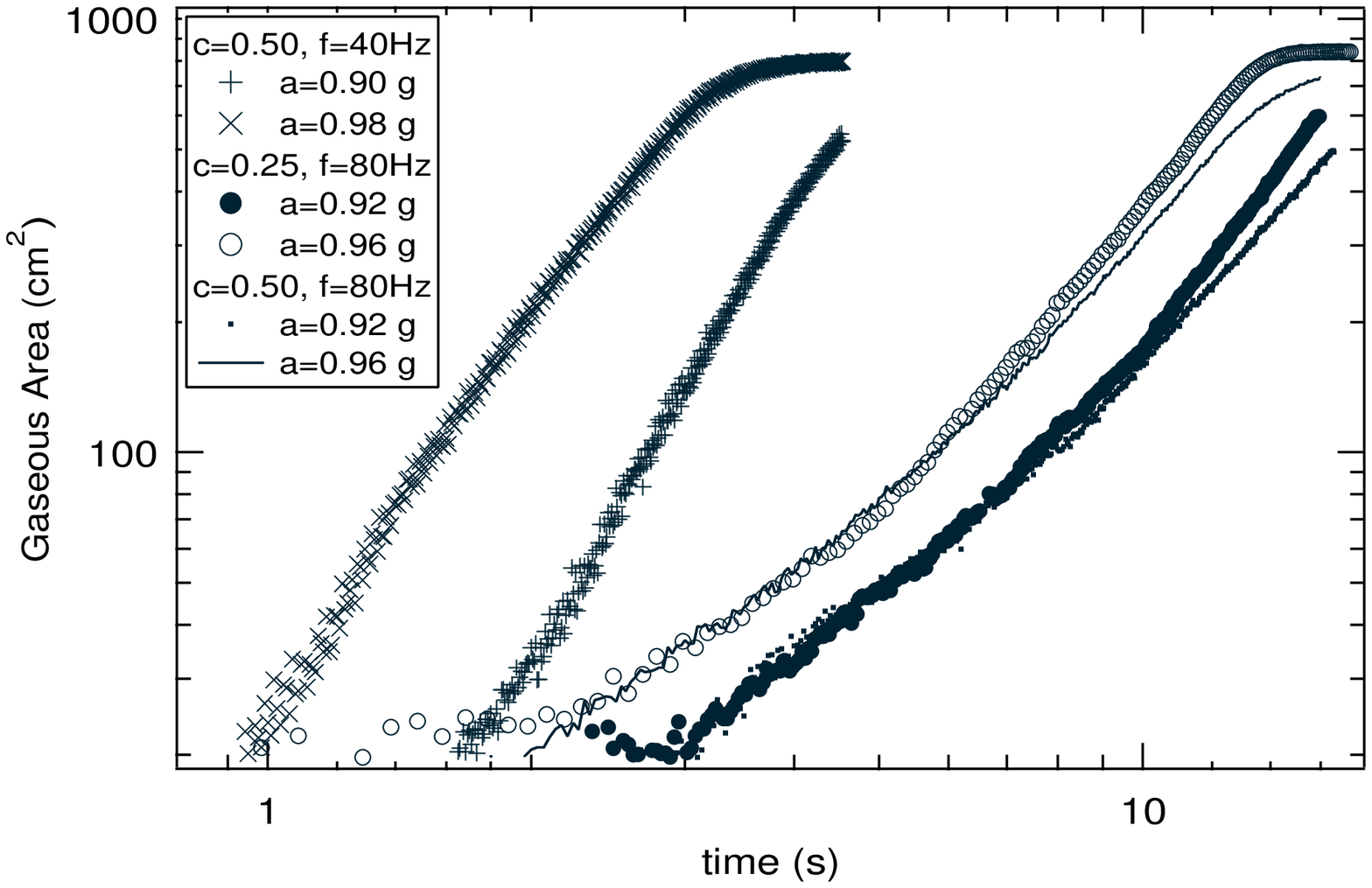, width=\linewidth} 
\end{center}
\caption{Growth of the gaseous area $A$ for various peak accelerations,
bead coverages, and vibration frequencies.
Smaller frequencies and larger accelerations lead to faster melting. 
Lower coverage leads to faster melting for large areas only.} 
\end{figure*} 

\begin{figure} 
\begin{center} 
\epsfig{file=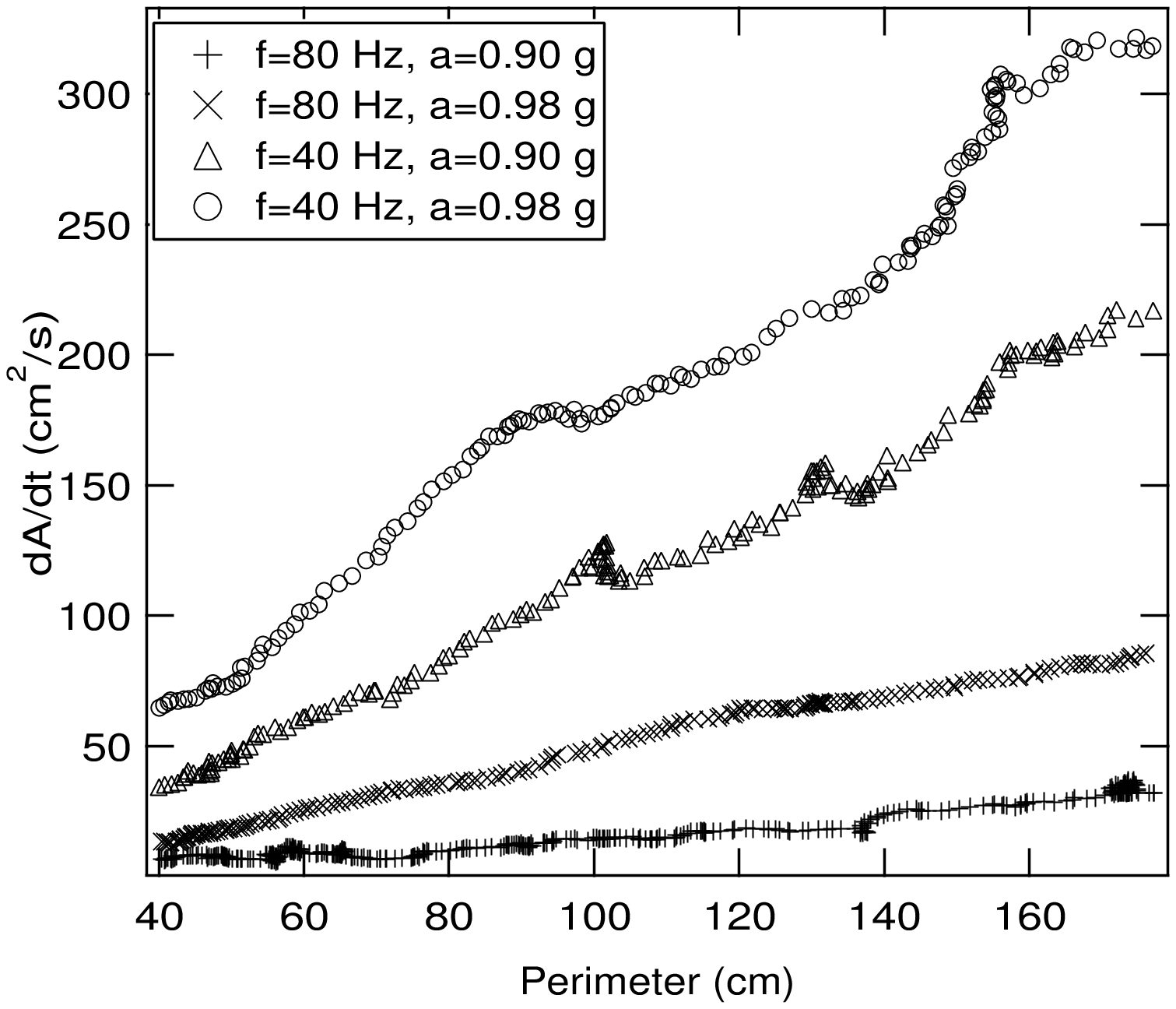, width=\linewidth} 
\end{center}
\caption{Growth rate of the gaseous area $dA/dt$ vs. perimeter length $P(A)$
for several peak accelerations and vibration frequencies.
The dependence is approximately linear.
Both $dA/dt$ and $P(A)$ are smoothed by means of 
a running average over 11 points.} 
\end{figure} 

\begin{figure} 
\begin{center} 
\epsfig{file=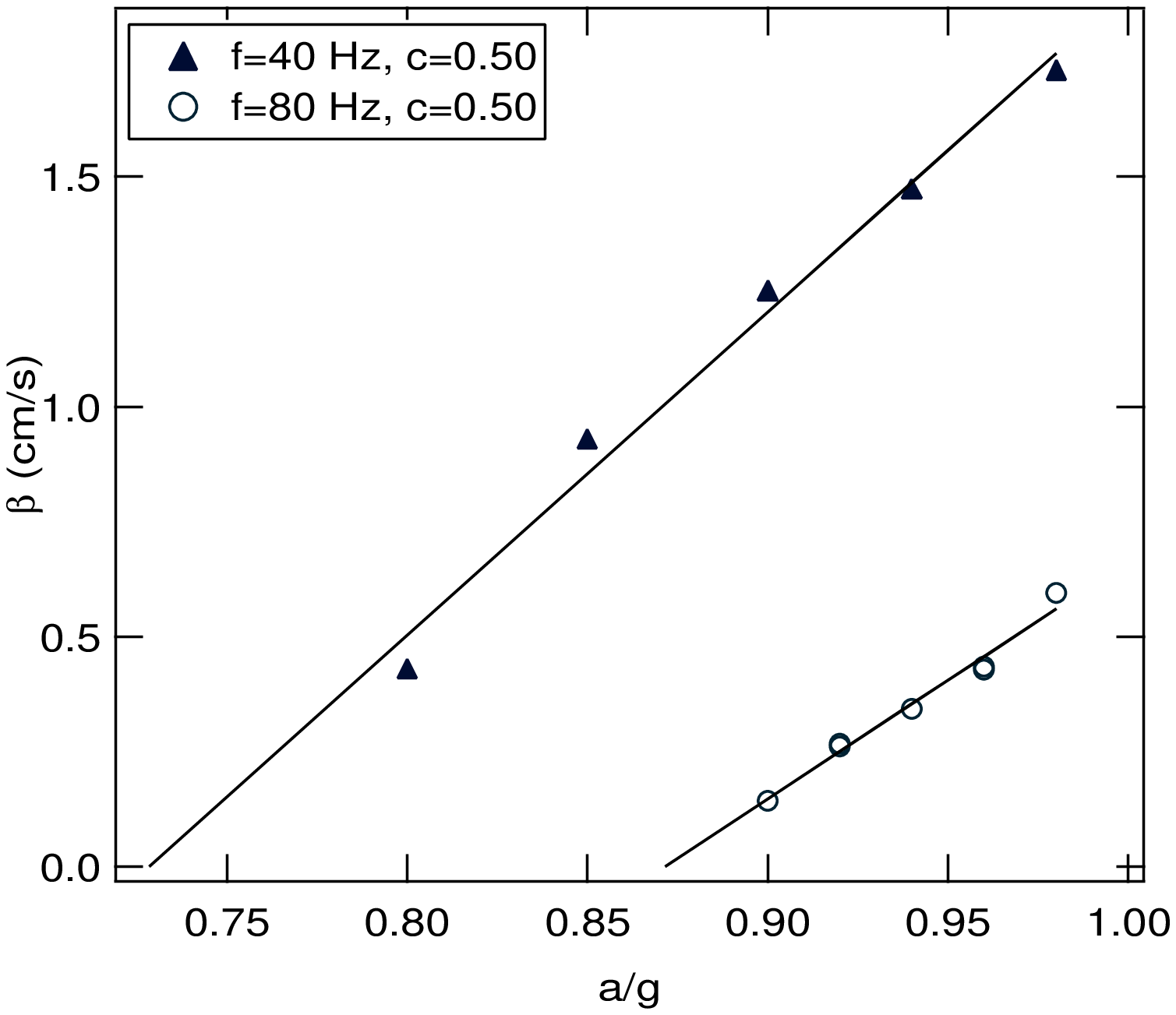, width=\linewidth} 
\end{center}
\caption{Front velocity $\beta$ vs. peak plate acceleration 
(normalized by g) for different
frequencies.
The dependence is linear; $\beta \to 0$ near $a_l$.} 
\end{figure} 

\begin{figure}
\begin{center} 
\epsfig{file=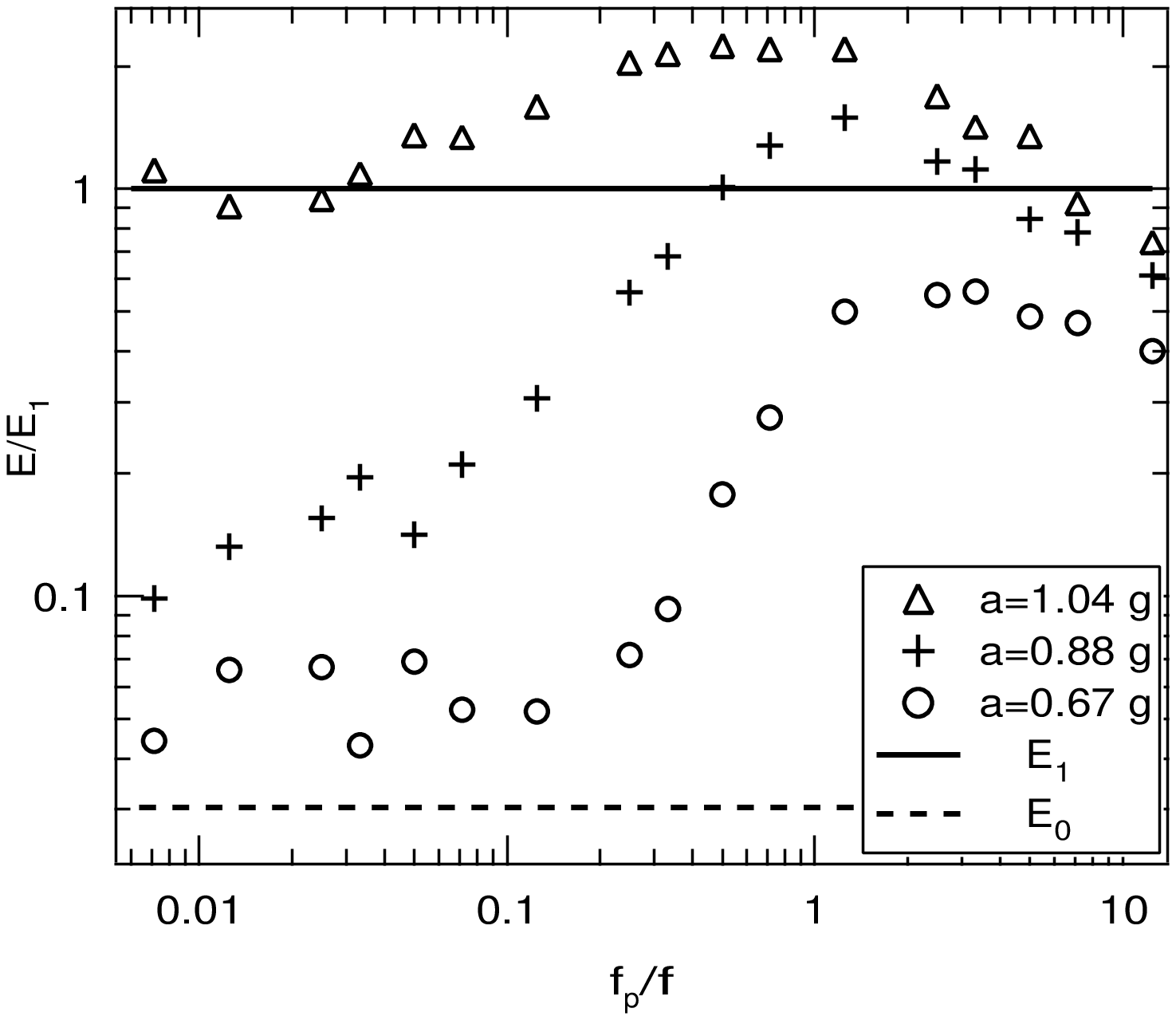, width=\linewidth} 
\end{center}
\caption{Simulation: Average energy of a bead vs. perturbation rate, and
comparison to energy of the lowest excited state $E_1$ and quiescent state
$E_0$ (horizontal lines). 
Above a threshold acceleration $a \approx 0.8$~g
the mean energy is larger than $E_1$ for a range of perturbation
rates, a necessary condition for maintenance of the excited state in 
the many particle system.} 
\end{figure} 

\begin{figure} 
\begin{center} 
\epsfig{file=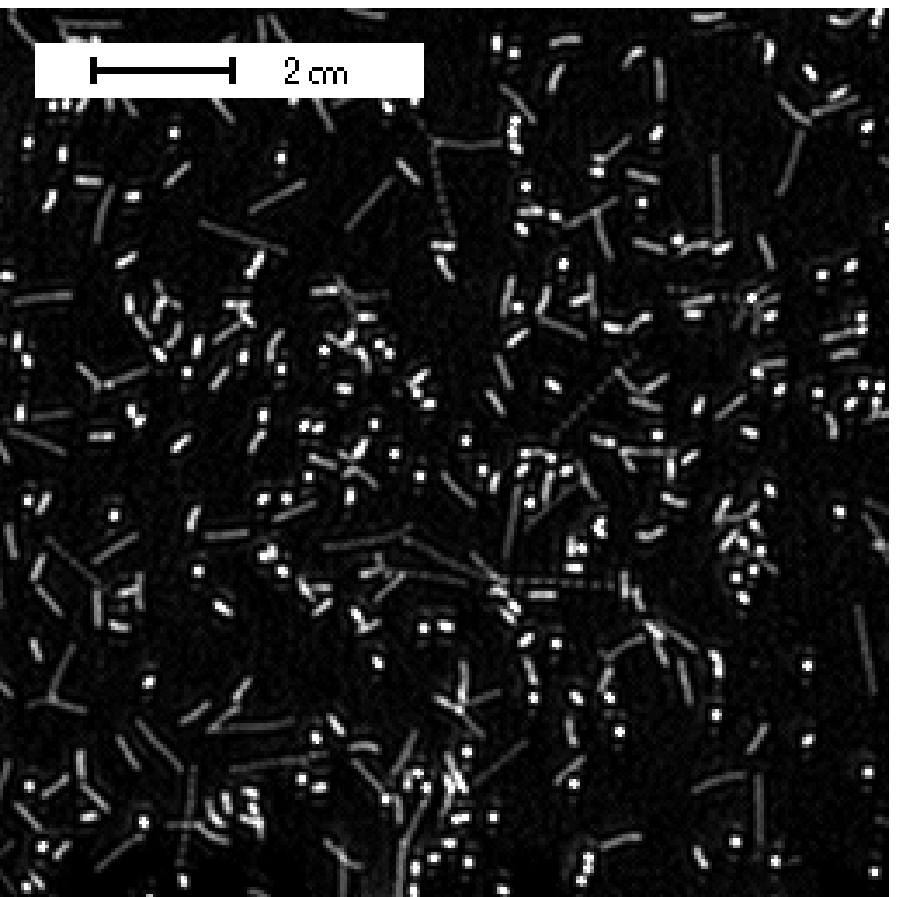, width=\linewidth} 
\end{center}
\caption{Coexistence of beads at rest and moving beads in a steady state
at low coverage.
Ten successive images, taken at $30$~frames/s were averaged
($c \approx 0.03$; $a=0.94 $~g; $f=80$~Hz).
Moving beads appear as streaks and stationary beads as bright spots.
Small clusters of beads at rest are clearly observable.} 
\end{figure} 


\begin{thebibliography}{99}
\bibitem{divcollisions} The numerical experiments see a divergence in the 
number of collisions during inelastic collapse. Experimentally
clustering has been observed, but the correspondence to 
numerical studies is unclear at this time.
\bibitem{foot1} For a more detailed description it may be preferable 
to treat the dense front as an additional transient phase rather than
considering it to be part of the gas.
\end{thebibliography}
\end{document}